\documentclass{elsart}
\usepackage{graphicx}

\usepackage{amssymb}

\def\etal{{\it et al.~}}
\def\eg{{\it e.g.,}}
\def\ie{{\it i.e.,~}}

\def\cm3{~{\rm cm^{-3}}}

\def\lsim{\mathrel{  
        \raise0.3ex\hbox{$<$}\kern-0.75em{\lower0.65ex\hbox{$\sim$}}}}
\def\gsim{\mathrel{
        \raise0.3ex\hbox{$>$}\kern-0.75em{\lower0.65ex\hbox{$\sim$}}}}

\begin{document}

\begin{frontmatter}



\title{Numerical Studies of Diffusive Shock Acceleration at Spherical Shocks}


\author[label1]{Hyesung Kang}
\ead{kang@uju.es.pusan.ac.kr}
\author[label2]{T. W. Jones},
\ead{twj@astro.umn.edu}
\ead[url]{www.astro.umn.edu/$\sim$twj}

\address[label1]{Pusan National University, Pusan 609-735, Korea}
\address[label2]{University of Minnesota, Minneapolis, MN 55455, USA}

\begin{abstract}
We have developed a cosmic ray (CR) shock code in one dimensional
spherical geometry with which the particle distribution, the gas flow
and their nonlinear interaction can be followed numerically in a frame 
comoving with an expanding shock.
In order to accommodate a very wide dynamic range of diffusion length scales 
in the CR shock problem, 
we have incorporated subzone shock tracking and adaptive mesh refinement techniques. 
We find the spatial grid resolution required for numerical convergence
is less stringent in this code compared to typical, fixed-grid Eulerian codes. 
The improved convergence behavior derives from maintaining the
shock discontinuity inside the same grid zone in the comoving
code. That feature improves numerical estimates of the compression rate
experienced by CRs crossing the subshock compared to codes that allow 
the subshock to drift on the grid.
Using this code with a Bohm-like diffusion model we have
calculated the CR acceleration and the nonlinear feedback at supernova
remnant shocks during the Sedov-Taylor stage.
Similarly to plane-parallel shocks, 
with an adopted thermal leakage injection model,
about $10^{-3}$ of the particles that pass through the shock
and up to 60 \% of the explosion energy are transferred to the CR component.
These results are in good agreement with previous nonlinear spherical
CR shock calculations of Berezhko and collaborators.
\end{abstract}

\end{frontmatter}

\section{Introduction}

Collisionless shocks form ubiquitously in tenuous cosmic plasmas via
collective interactions between particles and fields. 
The formation process of such shocks inevitably produces 
suprathermal particles, which can be further accelerated
to become cosmic rays (CRs hereafter) through the interactions with
resonantly scattering Alfv\'en waves in the converging flows across a shock
\cite{dru83,maldru01}. 
This so-called diffusive shock acceleration (DSA) is now widely
accepted as the primary mechanism through which nonthermal particles are
generated in a wide range of astrophysical environments. 
Both analytical calculations and numerical simulations have shown that DSA can be very efficient
and that there are substantial and highly nonlinear back reactions from 
the CRs to the bulk flows and to the MHD wave turbulence 
mediating the CR diffusive transport (\eg \cite{maldru01} and references therein). 

In the kinetic equation approach to numerical study of nonlinear CR acceleration at shocks,
the diffusion-convection equation for the particle momentum distribution,
$f(p)$, is solved along with suitably modified gasdynamic equations \cite{kj91}.
A variety of analytic and semi-analytic methods have been developed
to solve these equations for nonlinear, time-independent shocks (\eg 
\cite{mal98,blas02}). However, DSA modified shocks evolve as the
CRs within them extend to higher energies over time.
Many astrophysical shock systems
evolve dynamically on time scales not particularly long compared to
relevant DSA times, while the MHD waves that provide coupling
between the CRs and the bulk plasma also should evolve (\eg \cite{ptus03}). 
So, it is important to be able to model
time dependent, nonlinear DSA in such systems.
This will generally involve numerical simulations, a task that is quite
challenging, partly because
the full CR shock transition spans a very wide range of relevant length
scales associated with the particle diffusion lengths, $x_d(p)=\kappa(p)/u_s$,
from CR injection scales near the shock to outer diffusion scales for
the highest energy particles.
Here, $\kappa(p)$ is the momentum dependent spatial diffusion coefficient and
$u_s$ is the shock speed.
The full shock structure forms on a scale determined by the
highest momentum particles of dynamical significance; i.e.
comparable to $x_d(p_{max})$. The greater of $x_d(p_{max})$ or the
curvature radius
of the shock defines the size of the system to be computed. On the other hand
one must fully resolve flows on the scale of $x_d(p_{min})$ 
in order to follow properly the transport of the lowest momentum
CRs. That scale will generally be only somewhat greater than the
physical thickness of the viscous subshock that thermalizes
most of the plasma flowing through the shock.
These various scales may differ by many orders of magnitude for realistic
models of $\kappa(p)$.

Two approaches have so far been applied successfully to handle the
demanding spatial resolution issue in nonlinear, time dependent DSA simulations. 
Berezhko and collaborators (\eg \cite{berz94}) developed a method that
normalizes the spatial variable by $x_d(p)$ at each momentum
value of interest during solution of the CR kinetic equation. 
Their approach (which we here call `normalized grid' or NG for
convenience) allows efficient solution of the coupled system of time dependent
gasdynamic
equations and the CR transport equation even when the diffusion coefficient
has a strong momentum dependence (\eg $\kappa(p)\propto p$).
It does depend on knowing the diffusion coefficient {\it a priori}
of the simulations. 
The NG scheme was designed for simulations of supernova remnants, which have been 
represented by piston-driven spherical shocks in one-dimensional geometry.
Assuming Bohm-like diffusion, Berezhko and Ellison \cite{berel99,elber99} demonstrated qualitative agreement
between
nonlinear spherical DSA-modified shocks simulated with the above method
and steady, plane Monte Carlo shocks that allow escape of CRs
above a selected energy whose diffusion length corresponds
to the curvature radius of the shock. On the other hand, since
$\kappa(p)$ may evolve with the shock and may be nonuniform, one should have access to
alternative methods that allow inclusion of non-steady diffusion properties
and can still effectively handle the numerical challenges outlined
above.

As an alternative approach responding to these needs
Kang \etal developed the CRASH (Cosmic-Ray Amr SHock)
code in one dimensional (1D)
plane-parallel geometry by combining Adaptive Mesh Refinement
(AMR) techniques and a subgrid shock tracking technique \cite{kjls01}.
The effectiveness of AMR in this situation comes from the fact that the highest resolutions
are necessary only very close to the subshock, which
can still be treated as a numerical discontinuity satisfying standard Rankine-Hugoniot
relations, since the CR distribution is continuous across
this feature. Consequently, individual refinement patches can be made
quite small. By such efficient use of spatial gridding the blend of these 
computational strategies can greatly reduce the cost 
of time dependent DSA simulations, producing a practical tool
for modeling CR modified shocks with arbitrary diffusive
behavior. These benefits are especially valuable when combined with the 
so-called ``Coarse-grained
finite Momentum Volume'' (CGMV) method for solving the diffusion-convection
equation with a minimum of necessary information about the momentum
dependence of the CR distribution function \cite{jk05}.

NG calculations of supernova remnants by Berezhko and collaborators 
(\eg \cite{berz94,berz95,berz97,berz00})
have shown that, with Bohm-like diffusion,
up to 50 \% of explosion energy is converted to CRs when
an assumed fraction $10^{-4}-10^{-3}$ of incoming thermal particles are injected 
into the CR population at the subshock.
Nonlinear modification to the flow structure can be substantial.
Most obviously, as seen in many nonlinear CR shock simulations,
the total density compression through the full shock structure
is larger than the canonical gasdynamical value $(\gamma_g+1)/(\gamma_g-1)=4$
for strong gasdynamical shocks. In SNR simulations by Berezhko and V\"olk 
\cite{berz97,berz00}, for example, $ 4< \rho_2/\rho_0<15$.
Similar results obtain for nonlinear planar shocks. For example,
we previously applied our planar CRASH code 
to calculate the nonlinear evolution of CR modified shocks,
also assuming Bohm-like CR diffusion \cite{kj02,kj05}.
Rather than assume a fixed injection fraction at the subshock, 
those simulations adopted a more sophisticated``thermal leakage''
injection model that filters subshock crossings by suprathermal
particles using a momentum-dependent transparency function
based on a nonlinear plasma model for postshock wave-particle
interactions \cite{malvol98}. This injection model depends on 
a single, reasonably well-constrained parameter; namely, the ratio of the
upstream
magnetic field strength to downstream turbulent magnetic field strength. 
Indeed in our simulations $\sim 10^{-3}$ of incoming thermal particles were 
injected into the CR population at strong quasi-parallel CR modified shocks.
The ratio of CR energy to inflowing kinetic energy
increased with the shock Mach number, but 
approached $\approx 0.5$ for large shock Mach numbers, $M_s > 30$,
and it was relatively independent of other upstream properties or 
variation in the injection parameter.
The presence of a preexisting, upstream CR population is equivalent to having
slightly more efficient thermal leakage injection for such strong shocks,
while it can substantially increase the overall CR energy in moderate
strength shocks with $M_s<3$.

In the current work we discuss an extension of the CRASH code to 1D spherical geometry
intended to model CR acceleration in SNR and spherical wind shocks.  
In order to implement shock tracking and AMR techniques effectively
in this situation,
we solve the fluid and diffusion-convection equations in a frame 
comoving with the outer spherical shock. 
Thus, the shock and refined region around it stay at the same grid location
in the comoving frame, creating a relatively straightforward
scheme with all the advantages we found for plane geometry.
The basic equations and details of the numerical method are described
in \S 2.
We will present simulation results for a representative Sedov-Taylor blast
wave in \S 3, followed by a summary and comparison 
to previous simulation results in \S 4.
Future papers will apply this code to the study of SNRs and their
emissions as well as to quasi-spherical supersonic astrophysical winds.

\section{Numerical Method}
\subsection{Basic Equations}
The evolution of CR modified shocks depends on a coupling between
the gasdynamics and the CRs. That coupling takes place by way of
resonant MHD waves, although it is common practice to express
the pondermotive wave force and dissipation in the plasma
using the associated CR pressure distribution properties along with
a characteristic wave propagation speed (usually the Alfv\'en speed).
Consequently, we solve the standard gasdynamic equations with CR pressure terms
added in the conservative, Eulerian
formulation for one dimensional spherically symmetric geometry. 
For strongly shocked flows
numerical errors in computing the gas pressure from the total
energy can lead to spurious entropy generation with
standard methods, especially in the shock precursor.
To avoid this problem we replace the usual energy conservation
relation with the
evolution of a modified entropy, $S = P_g/\rho^{\gamma_g - 1}$, 
everywhere except across the subshock \cite{kjg02}.
Energy conservation is applied across the subshock.
The resulting dynamical equations are:

\begin{equation}
{\partial \rho \over \partial t}  +  {\partial\over \partial r} (\rho u) 
= -{ 2 \over r} \rho u,
\label{masscon}
\end{equation}

\begin{equation}
{\partial (\rho u) \over \partial t}  +  {\partial\over \partial r} (\rho u^2 + P_g + P_c) 
= -{2 \over r} \rho u^2,
\label{mocon}
\end{equation}

\begin{equation}
{\partial (\rho e_g) \over \partial t} + {\partial \over \partial r} 
(\rho e_g u + P_g u) = 
-u {{\partial P_c}\over {\partial r}} 
-{2 \over r} (\rho e_g u + P_g u), 
\label{econ}
\end{equation}

\begin{equation}
{\partial S\over \partial t}  +  {\partial\over \partial r} (S u) =
-{2\over r} S u
+ {(\gamma_{\rm g} -1)\over \rho^{\gamma_{\rm g} -1} } [W(r,t) - L(r,t)], 
\label{scon}
\end{equation}

where $P_{\rm g}$ and $P_{\rm c}$ are the gas and the CR pressure,
respectively, $e_{\rm g} = {P_{\rm g}}/{[\rho(\gamma_{\rm g}-1)]}+ u^2/2$
is the total energy of the gas per unit mass.
The remaining variables, except for $L$ and $W$ have standard meanings.
The injection energy loss term, $L(r,t)$, accounts for the
energy carried by the suprathermal particles injected into the CR component at
the subshock and is subtracted from the postshock gas 
immediately behind the subshock. 
Gas heating due to Alfv'en wave dissipation in the upstream region is 
represented by the term
\begin{equation}
W(r,t)= - v_A {\partial P_c \over \partial r }, 
\end{equation}
where $v_A= B/\sqrt{4\pi \rho}$ is the Alfv'en speed.
This term derives from a simple model in which Alfv'en waves are amplified by
streaming CRs and dissipated locally as heat in the precursor region
(\eg \cite{mckenzi82,jon93}).

The CR population is evolved by solving the diffusion-convection equation ,
\begin{equation}
{\partial g\over \partial t}  + (u+u_w) {\partial g \over \partial r}
= {1\over{3r^2}} {\partial \over \partial r} [r^2 (u+u_w)]( {\partial g\over
\partial y} -4g) + {1 \over r^2}{\partial \over \partial r} [r^2 \kappa(r,y)  
{\partial g \over \partial r}], 
\label{diffcon}
\end{equation}
where $g=p^4f$, with $f(p,r,t)$ the pitch angle averaged CR 
distribution, 
and where $y=\ln(p)$, while $\kappa(r,p)$ is the diffusion coefficient
\cite{skill75}.
For simplicity we always express the particle momentum, $p$ in
units $m_{\rm p}c$ 
and consider only the proton CR component.
The wave speed is set to be $u_w=v_A$ in the upstream region, while we
use $u_w=0$ in the downstream region.
This term reflects the fact that
the scattering by Alfv'en waves tends to isotropize 
the CR distribution in the wave frame rather than the gas frame 
\cite{skill75}. Upstream, the waves are expected to be dominated by the
streaming instability, so face upwind. Behind
the shock various processes, including wave reflection, are
expected to lead to a more nearly isotropic wave field (\eg \cite{achbl86}).

\subsection{The CRASH code on a Comoving Spherical Grid}  

Before introducing the spherical version of the CRASH code, it is
helpful to outline the basic strategy of the analogous planar code.
The full CR shock transition involves
a very wide range of length scales 
associated with the particle diffusion lengths that must be resolved to 
follow the shock evolution properly. 
In particular, structure must be resolved down close to the subshock 
physical thickness, since the diffusion scales for low energy, freshly
injected particles are expected to be comparable.
The outer scale of any DSA calculation is set by the greater of the
full width of the shock transition or the
size of the physical system being studied. 
Although it is necessary to include very fine scales to resolve interactions of
the lowest energy CRs, their diffusion and acceleration are
important over distances only within a few times $x_d(p_{inj})$ of the subshock,
where $p_{inj}$ represents a characteristic injection momentum.
Grid refinement in the CRASH code, therefore, 
includes a region spanning the subshock only large enough to include 
comfortably the diffusion scales of dynamically important CRs and 
enough levels to follow freshly injected CRs adequately \cite{kjls01}.
In CRASH each level of refinement has twice the spatial resolution
of the level below, but the same number of grid zones.
Typically, we have used about 200 zones in each refinement grid.
To accomplish grid refinement effectively it is necessary to locate
the subshock position exactly. Thus, we track the subshock
as a moving, discontinuous jump inside the initial, uniform and
fixed grid \cite{kjls01}.
The subshock jump conditions and motion of the subshock
are established by solution of the nonlinear gasdynamic Riemann problem.
An additional measure is used in the planar CRASH code to ensure that the 
shock remains near the middle of the refined region at all grid levels 
during each base-grid time step.  In particular
the fluid velocity in the refinement patch is transformed to the subshock rest frame 
calculated at the start of each base-grid time step.
All the refined grids are then accordingly redefined at each time 
step.
Thus, in the planar CRASH code the refinement patch moves smoothly with the shock,
while the base grid is fixed in space.
This greatly simplifies the refinement strategy, especially since it
eliminates the need for construction of new refinement
patches after the simulation is started. 

A somewhat different grid strategy is needed in spherical geometry, 
since the analogous Galilean transformation cannot be applied.
Drury and Mendonca \cite{drury00} pointed out that a spherical shock 
can be made to be stationary by adopting comoving variables 
which factor out a uniform expansion or contraction. 
Here we define a comoving frame that expands with the 
instantaneous shock speed. This strategy for simulations of
SNR flows has, of course, been applied in various forms
before (\eg \cite{chev95,jun96}) and 
is the standard approach in cosmological simulations
(\eg \cite{ryu93,martel98}). 
Following the conventional cosmological formalism, 
we adopt the {\it comoving} radial coordinate, $x=r/a$, 
where $a$ is the expansion factor and $a=1$ at the start of simulations.
The expansion rate, $\dot a = (u_s - v_s)/x_s$, is found from the
condition that the shock speed is zero at the comoving frame.
Here $u_s$ and $v_s$ are the shock radial velocities in the Eulerian frame
and in the comoving frame, respectively.
Then the {\it comoving} density and pressures are defined as $\tilde \rho = \rho a^3$,
$\tilde P_g = P_g a^3$, and $\tilde P_c = P_c a^3$.
The gasdynamic equation with CR pressure terms in the spherical comoving frame
can be written as follows:

\begin{equation}
{\partial \tilde \rho \over \partial t}  + {1\over a} {\partial \over \partial x
} (\tilde \rho v)
= -{ 2 \over {ax} } \tilde \rho v,
\label{masscon2}
\end{equation}

\begin{equation}
{\partial (\tilde \rho v) \over \partial t}  +  {1 \over a} {\partial \over 
\partial x}
(\tilde \rho v^2 + \tilde P_g + \tilde P_c)
= -{2 \over {ax}} \tilde \rho v^2 - { \dot a \over a}\tilde \rho v
- \ddot a x \tilde \rho,
\label{mocon2}
\end{equation}

\begin{equation}
{\partial (\tilde \rho \tilde e_g) \over \partial t}  + {1 \over a} {\partial
\over \partial x} (\tilde \rho \tilde e_g v + \tilde P_g v )= 
- {v \over a} {{\partial \tilde P_c}\over {\partial x}}
-{2 \over {ax}} (\tilde \rho \tilde e_g v + \tilde P_g v) - 2{\dot a \over a}\tilde \rho 
\tilde e_g - \ddot a x \tilde \rho v, 
\label{econ2}
\end{equation}

\begin{equation}
{\partial \tilde S\over \partial t}  +  {1 \over a} {\partial \over \partial x}
 (\tilde S v) =
-{2\over {ax}} \tilde S v - 2 {\dot a \over a} \tilde S
+ {(\gamma_{\rm g} -1)\over \tilde \rho^{\gamma_{\rm g} -1} } [\tilde W(x,t) - \tilde L(x,t)]. 
\end{equation}
As mentioned before, the entropy Eq. (10) is solved everywhere outside 
the subshock, while the energy conservation Eq. (9) is applied across 
the subshock.
The deceleration rate is calculated numerically at a give time step  by
$\ddot a = (\dot a^n - \dot a^{n-1})/\Delta t^n$,
where $\dot a^{n-1}$ is the expansion rate at the previous time step.
The diffusion-convection equation for the function
$\tilde g=p^4 \tilde f$, is given by
\begin{eqnarray*}
\lefteqn{ {\partial \tilde g\over \partial t}  + {(v+u_w)\over a} {\partial \tilde g \over 
\partial x}= }
\end{eqnarray*}
\begin{equation}
\lefteqn{
[{1\over{3ax^2}} {\partial \over \partial x}[x^2(v+u_w)] + {\dot a \over a} ]
( {\partial \tilde g\over
\partial y} -4\tilde g) + 3{\dot a \over a}\tilde g + {1 \over {a^2x^2}}
{\partial \over \partial x} [x^2
\kappa(x,y)  {\partial \tilde g \over \partial x}],}
\label{diffcon2}
\end{equation}
where $\tilde f(p,r,t)$ is the {\it comoving}
CR distribution function.
This equation automatically includes adiabatic losses for
the CRs resulting from the spherical expansion of the flow.
The equations \ref{masscon2} - \ref{diffcon2} are solved
using finite difference methods analogous to those outlined
for the planar CRASH code and discussed more extensively
in \cite{kjls01}. 

\subsection{The Thermal Leakage Injection Model}

In the ``thermal leakage'' model for CR injection at shocks, most of
the downstream thermal protons are locally confined by nonlinear MHD waves
even when their speed exceeds the speed difference between the shock
and the bulk downstream plasma. 
Only particles well into the tail of the postshock Maxwellian distribution 
can leak upstream across the subshock
\cite{malvol98,maldru01}.
In particular, leaking particles not only must have velocities that would
allow them to reach the receding subshock,
they must also avoid being scattered by the MHD waves that mediate the plasma subshock. 
This filtering process is implemented numerically in the CRASH code by
adopting a ``transparency function'', $\tau_{\rm esc}(\epsilon_B, \upsilon)$,
that expresses the probability of supra-thermal
particles at a given velocity, $\upsilon$, leaking upstream through the 
postshock MHD waves \cite{gies00,kjg02}.
One free parameter controls this function; namely, $\epsilon_B = B_0/B_{\perp}$,
which is the inverse ratio of the amplitude
of the postshock MHD wave turbulence $B_{\perp}$ to the general magnetic field
aligned with the shock normal, $B_0$ \cite{malvol98}.
Plasma hybrid simulations and theory both suggest that
$0.25 \lesssim \epsilon_B \lesssim 0.35$ \cite{malvol98},
so that the model is well constrained. 
We have also found for strong shocks that the time-asymptotic
behaviors are only weakly dependent on the choice of this parameter.
In this model
the subshock is completely ``opaque'' to downstream
particles with momenta less than $p_1$,
({\it i.e.,} $\tau_{esc}=0$ for $p<p_1$,
where $p_1 = (u_2/c) (1+\epsilon_B)/\epsilon_B$).
However, the subshock becomes virtually transparent
to particles with $p> (2-3) p_1$ (\ie $\tau_{esc} \rightarrow 1$). 
We have ported this injection scheme
to the spherical CRASH code, and will apply it to the simulations
presented below.

\subsection{A Bohm-like Diffusion Model}
DSA, along with the time and length scales for CR acceleration,
and subsequent shock modification, depend on the spatial
diffusion coefficient for the CRs.
The diffusion coefficient can be expressed
in terms of a mean scattering length, $\lambda$, as
$\kappa(x,p) = {1\over 3} \lambda \upsilon$. 
The Bohm diffusion model is commonly used to represent a saturated 
wave spectrum, providing
the minimum expected diffusion coefficient in a quasi-parallel shock.
We adopt a variation on that model in the simulations presented here.
Bohm diffusion assumes that CRs scatter within one gyration 
radius (\ie $\lambda \sim r_{\rm g}$).
This gives $\kappa_{\rm B} \propto ~{p^2}/{(p^2+1)^{1/2}}$
with $\kappa_{\rm B} \propto p^2$ in the limit of $p<<1$ and 
$\kappa_{\rm B} \propto p$ in the limit of $p>>1$.
Because the nonrelativistic momentum dependence is very steep, 
the diffusion coefficient and, thus, the diffusion length
of freshly injected CRs, is very much smaller than for relativistic
CRs if the shock speed is a small fraction of the speed of light. 
Since it is necessary, at least near the subshock, to
resolve all relevant diffusion lengths, use of a Bohm diffusion model
requires extremely fine grid resolution where freshly
injected CRs are concentrated.
On the other hand, previous calculations have shown that,
although details in diffusion models at the lowest momenta
influence the thermal leakage injection rate and
early shock evolution, they do not play significant roles
once at least mildly relativistic particles dominate the CR pressure
(\eg \cite{kjls01}). 
Consequently, we will apply in the simulations
presented here the following Bohm-like diffusion coefficient
that includes a weaker nonrelativistic momentum dependence; namely,
\begin{equation}
\kappa(\rho,p) = \kappa_{\rm n} p \left(\frac{\rho_0}{\rho}\right).
\end{equation}
The coefficient 
$\kappa_{\rm n} = m c^3/(3eB) = (3.13\times 10^{22} {\rm cm^2s^{-1}} ) B_{\mu}^{-1}$,
where $B_{\mu}$ is the magnetic field strength in units of microgauss. 
The assumed density dependence for $\kappa$ accounts for compression of the
perpendicular component of the wave magnetic field and also
inhibits the acoustic instability in the precursor of highly modified
CR shocks \cite{kjr92}.
We note, also, that hereafter we use the subscripts '0', '1', and '2' to denote
conditions far upstream of the shock, immediately upstream of the
gas subshock and immediately downstream of the subshock, respectively.
Thus, $\rho_0$ represents the far-upstream gas density,
which will be constant in the simulations presented here.

\section{Simulations of Sedov-Taylor Blast Waves}

For a supernova remnant (SNR) propagating into a uniform ISM (interstellar
medium)
the CR acceleration takes place mostly during free expansion
and Sedov-Taylor (ST hereafter) stages,
since the shock slows down significantly afterward.
In fact the highest momentum, $p_{\max} \sim 10^5$, which corresponds to
the proton energy of $10^{14}$eV for a typical SNR, is roughly achieved
by the end of the free expansion stage, while the transfer of explosion
energy ($E_o$) to the CR component occurs mostly during the ST stage.
This is because the total CR energy gain is proportional to the kinetic
energy passed through the shock, $E_{sw}= 2\pi \int \rho_0 u_s^3 r_s^2 dt$, 
and $E_{sw}$ becomes comparable to $E_o$ by the end of ST stage.
In this paper, we calculate the CR acceleration during this stage,
since the current version of CRASH code treats only the forward shock.
Application of our new AMR algorithm for multiple spherical shocks is
not simple, since it requires multiple, comoving grids.
Although CR acceleration at the reverse shock 
can be dynamically important during the free expansion stage \cite{ellison05}, 
CR energy generation occurs mostly at the forward shock during ST stage.
Adiabatic losses by CRs accelerated early on in the interior and then
advected outward through the ST phase would generally be very large.

\subsection{Example SNR Model Parameters}
For most of the simulations we consider a supernova explosion with $E_o=10^{51}$ ergs and $M_{sn}=10
M_{Sun}$ in a warm ISM with a uniform density,  
$n_{\rm H}=0.3 {\rm cm}^{-3}$ and $T=10^4$ K.
The physical quantities are normalized, both in the numerical code and in
the plots below, by the following constants:
$\rho_o = 7.0\times 10^{-25} {\rm g~cm}^{-3}$,
$t_o=1.3\times 10^3$ yr, $r_o=6.14$ pc, $u_o=4.6\times 10^3 {\rm km~s}^{-1}$,
and $P_o= 1.5\times 10^{-7} {\rm erg ~cm^{-3}}$. 
We assume a Bohm-like diffusion coefficient,
$\kappa= (6.26\times 10^{21} {\rm cm^2~s^{-1}}) p (\rho_0/\rho)$,
where $B_{\mu}=5$ is adopted for the mean ISM magnetic field.
The Alfv'en speed is given by $v_A= v_{A,0}/(\rho/\rho_0)^{1/2}$  
where $v_{A,0} = 17 {~\rm km s^{-1}}$ is for the background ISM.
The pressure of the background gas is set to be
$P_{g,0} = 10^{-12}{\rm erg ~cm^{-3}}$ 
so the gas sonic Mach number ($M_s= u_s/\sqrt{\gamma_g P_{g,0}/\rho_0}$)
of the initial shock is $M_s=130$. 
In order to explore effects of pre-existing CRs, 
we include an ambient (upstream) CR population, 
$f(p)\propto p^{-4.5}$, and set its pressure as either
$P_{c,0}=0.005P_{g,0}$
or $P_{c,0}=0.5P_{g,0}$.
In fact, the injection is so efficient for the strong shocks considered
here that the presence of these pre-existing CRs is effectively
equivalent to having an only slightly higher injection rate. It
does not qualitatively affect the energy extracted by CRs from
the shock or the degree of shock modification.

Although the theoretically preferred values of $\epsilon_B$ lie between
0.25 and 0.35 for strong shocks \cite{mal98},
values as large a these lead to very efficient initial thermal leakage 
injection for shock
Mach numbers greater than 100 and a rate of shock evolution  at
simulation start-up
that is very difficult to follow numerically. 
To avoid those start-up issues, we have to adopt a smaller value 
for the current simulations.
Also, with the assumed $\kappa$, thermal leakage injection of suprathermal
particles is more efficient, compared to the true Bohm diffusion case.
So we adopted $\epsilon_B\sim 0.16$
in order to obtain a typical injection fraction of $\sim 10^{-3}$.
Previously we showed that the CR acceleration depend only weakly
on the values of $\epsilon_B$ \cite{kjls01}, so this choice will
not influence our conclusions.

For the simulations presented here we use 230 uniformly spaced 
logarithmic momentum zones in the interval $y=\ln p = [\ln p_1,\ln p_2]$
when solving Eq. (11), where $p_1$ is determined 
instantaneously by the downstream flow speed, $u_2$ and 
$ \ln p_2 = 15.4$. As the postshock density compression increases,
$u_2$ and $p_1$ decrease in time, so the particle distribution 
function $g(y)$ from the previous time step solution is mapped onto the new 
momentum zones.

The simulations are initialized at $\hat t = (t/t_o)=1$ by the ST similarity 
solution, which is characterized by the shock position and speed expressed in
code units,
$ \hat r_s = (r_s/r_o) = \xi_s \hat t^{2/5}$ and $\hat u_s = (u_s/u_o)= 
(2/5)\xi_s \hat t^{-3/5}$ 
with $\xi_s=1.15167$.  The spatial resolution in the
base grid is 
$\Delta \hat r_0 = 6.0 \times 10^{-4}$, expressed in code units.
In the {\it 8th} refined grid, which is the finest we have used
in the simulations presented here, the resolution is
$\Delta \hat r_8 = 2.3 \times 10^{-6}$.
Remarkably, we found that although $\Delta \hat r_8$ is much larger than the diffusion length for
$p_1 \approx 10^{-2}$, $\hat l_{diff}= 2.5\times 10^{-9}$,
the spherical, comoving CRASH simulations show good numerical convergence. 
In particular, we found that an increase from 8 to 10 levels of refined grids,
corresponding to a factor 4 improvement in spatial resolution,
leads to an increase of $P_c$ by less than 0.5 \% (Convergence
in $P_c$ is generally from below with our scheme.) Other variables
respond similarly.
This behavior is contrary to what we found in comparison simulations using
a fixed, Eulerian
grid. There, in agreement with our previous experience using
Cartesian grids, spatial resolution several times finer than the diffusion length associated with $p_1$ 
were necessary for adequate convergence \cite{kjls01}. 
This will be discussed further below.

\subsection{Basic Code Tests}

As an initial test of the spherical CRASH code, we calculated 
identical ST blast waves on both a fixed Eulerian grid and a comoving grid.
For the fixed Eulerian grid simulation our AMR scheme cannot be applied
simply, because
it requires that the refined patch move with the shock. Consequently,
only the base grid ($l_g=0$) with $\Delta \hat r_0 = 6.0 \times 10^{-4}$ 
is used in that simulation. Without AMR it is not practical to carry out that
simulation with the high resolutions applied in subsequent AMR tests.
We used instead the same base grid applied for the other simulations, 
but adopted a large diffusion coefficient,
$\tilde \kappa = 0.1~ p(\rho_0/\rho)$, to maintain adequate numerical convergence.
In the comoving grid version of the test, two grid structures are considered; namely,
the base grid alone, without AMR (\ie the same as the fixed grid simulation except for the expansion),
and the base grid with a single ($l_g=1$) refinement level using $\Delta \hat r_1 = 3.0 \times 10^{-4}$. 
The large diffusion coefficient in this test model leads to very slow CR acceleration
and associated shock modification. 
In order to introduce at least modest CR feedback effects
the initial shock was much weaker than we used in follow-up
tests, and the ambient CRs dominated the upstream pressure;
namely, $P_{c,0}= 4 P_{g,0}$, with $f(p)\propto p^{-4.5}$.
The thermal leakage injection and Alfv\'en wave terms were turned off for this
test. 
We assumed a hot ISM in which $n_H=3\times 10^{-3} {\rm cm}^{-3}$ 
and $T=10^6$ K with the initial shock Mach number $M_s=13$.
The normalization constants are:
$\rho_o = 7.0\times 10^{-27} {\rm g~cm}^{-3}$,
$t_o=6.1\times 10^3$ yr, $r_o=28.5$ pc, $u_o=4.6\times 10^3 {\rm km~s}^{-1}$,
and $P_o= 1.5\times 10^{-9} {\rm erg ~cm^{-3}}$.
Fig. 1 shows that the three test simulations with different grid choices
are in good agreement, demonstrating the validity of the comoving grid
method. 

Our second test explored the convergence behavior of the spherical
CRASH code. The physical problem was the same as for the first test,
except that we used a much smaller and physically
more interesting diffusion coefficient,
$\tilde \kappa = 1.5\times 10^{-7}~p(\rho_0/\rho)$
and a smaller ambient CR pressure, $P_{c,0} = 0.5 P_{g,0}$.
As a benchmark, one simulation was carried out on a uniform, fixed grid identical
to that of the previous test. 
The CRASH, comoving simulations were done on a uniform grid both without
AMR, $l_g=0$, and with AMR using $l_g=1, 3, 5, 8, {\rm~and}~ 10$.
Results are shown in Fig. 2.
The AMR simulation results converge well as the number of refinement
levels increases.
CR acceleration efficiency increases with finer resolution, consistent
with previous results (\eg \cite{kjls01}). 
As expected, the coarse fixed grid simulation and the comoving simulations
with coarser resolution are not very accurate as measured by the
high resolution solutions. It is interesting to note, however, that
the CR acceleration is more accurate in the CRASH simulation 
with the comoving grid (with $l_g=0$) than the
fixed Eulerian grid simulation of comparable resolution. 
Thus, the comoving grid provides a more economical solution to
the problem than the fixed grid.
This can be understood as follows. In the comoving grid simulation,
the peak of the spatial distribution of $f(p)$ and the subshock location remain
in the same grid zone throughout the calculation. 
Consequently, the subshock compression ($\nabla \cdot \vec u$),
which determines the rate at which low energy CRs accelerate (see Eq. [11]),
is applied to the CR distribution consistently.
In the fixed Eulerian grid, on the other hand, the shock drifts through
the grid as the shock evolves. As this happens the numerical
estimate of subshock compression varies and, on average is reduced.
This reduces the effective adiabatic compression applied to $f(p)$
at the subshock.
This behavior also leads to the the unsteady behavior of the CR pressure at the subshock,
$P_{c,2}$, for the fixed grid simulation, shown  in the lower right 
panel of Fig. 2.

\subsection{Results}
Figs. 3 and 4 show the early evolution of the ST blast wave outlined
in \S 3.1 (initial Mach number 130) over the interval $1.0 \le \hat t \le 1.5$.
The simulation shown in Fig. 3 includes the Alfv\'en wave
terms, while the simulation shown in Fig. 4 turns them off. Otherwise
they are identical.
The black solid lines show the initial similarity solution 
with $P_{c,0}=0.5P_{g,0}$ at $\hat t=1$.
The CR pressure reaches the highest value during the initial 130 years 
($\Delta \hat t = 0.1$), but it decreases as the shock slows down over time .
As previously shown in SNR simulations by Berezhko et al. \cite{berz97},
and anticipated from earlier work (\eg \cite{mckenzi82,ach82,mark92,jon93}),
the simulation with Alfv'en wave drift and heating 
terms (Fig. 3) shows less efficient CR acceleration than
the simulation without those terms (Fig. 4).
After 650 years ($\Delta \hat t=0.5 $), $p_{max} \approx 10^4$,  as indicated
by the exponential cut-off in the particle distribution function
$g(p)$ at the shock shown in the lower left panel of Figs. 3 and 4.
In both simulations, nonlinear modifications due to CRs are significant,
represented clearly by growth of a significant precursor and concave curvatures
in the CR particle distribution. 

The density in the precursor just in front of the subshock, $\rho_1$, and in the
postshock region, $\rho_2$, immediately after the subshock,
respectively, are shown in the top panels of Fig. 5.
In the model with the Alfv\'en wave terms, the precursor compression stays around 2,
while the total compression stays around 7-8.
Without the wave terms, on the other hand,
the CR acceleration is very efficient and
the total density compression can go up to a factor of 30.
This last result is consistent with our simulations of plane-parallel
shocks omitting wave terms, 
\ie $\rho_2/\rho_0 \sim 1.5 M_{s,0}^{0.6}$ where $M_{s,0}$ is the Mach number of
a unmodified shock \cite{kjg02} as well as previous
SNR simulations without these terms (\eg \cite{berz95}). 
The second from the top panels show the Mach number of the 
subshock, which decreases quickly during the initial phase
and stays around $M_s\approx 5$ for the model with the wave terms.
The third from the top panels show the 
the CR pressure and gas pressure normalized to the ram pressure of 
the unmodified ST similarity solution, 
\ie $\rho_0 u_{ST}^2 \propto \hat t^{-6/5}$.
After the initial quick changes, these ratios remain constant at
$P_{c,2}/(\rho_0 u_{ST}^2) \approx 0.4$ and
$P_{g,2}/(\rho_0 u_{ST}^2) \approx 0.2$ in the model with the wave terms.
Consequently, the conversion factor from the shock kinetic energy to the CR energy
approaches an asymptotic value, although the pressure itself decreases
as the remnant expands.
The bottom panels show the fraction of particles that have passed through
the shock and injected into the CR component.
With the adopted value of $\epsilon_B=0.16$, this fraction is 
$\xi \sim 2 \times 10^{-4} - 2 \times 10^{-3} $ 
in the model with the wave terms,
while it can be quite large, $\xi \sim 0.01$,  during the 
initial phase in the model without the wave terms. 

Fig. 6 shows the total CR numbers integrated over the simulated volume, 
$N_p=  \int r^2 dr  f(p) p^2 $, and
$G_p = N_p p^2 = \int r^2 dr g(p)$ in code units. 
The slope of the integrated spectrum, $q = - d (\ln N_p)/ d \ln p $,
is shown in the bottom panels.
We note that it is much easier to recognize the nonlinear concave
curvature in the plot of $G_p$, while $N_p$ resembles
a power law below $p_{max}$.
In the model with wave terms,
the slope for nonrelativistic particles is $q \approx 2.2$,
consistent with the compression ratio across the subshock with
$M_s\approx 4.6$. 
The slope flattens to $q\sim 1.6$ near the highest momentum,
above which the spectrum decreases exponentially.
This agrees well with Berezhko and V\"olk \cite{berz97} where
the power law index of $N_p$ is about 1.7 for $p>10^3$, and
steeper at lower energy for the model with their injection rate, 
$\eta = 10^{-3}$. The steep momentum slope at low energies results 
in nonlinear CR shocks from
the fact that these CRs are accelerated primarily through localized
crossings of the subshock. The flat high energy slope, on the
other hand, reflects the fact that these CRs, with their large
mean free scattering paths, are accelerated
across the full shock precursor, so the relevant compression is greater.

Finally, we show in Fig. 7 the total kinetic, thermal and CR energies integrated over the simulation
volume, $E_{kin}$, $E_{th}$, $E_C$, 
respectively.  This shows that the fraction of total kinetic energy 
remains the same as that of the initial similarity solution. 
But, by $\hat t = 10$ the gas thermal energy fraction decreases 
as the CR energy fraction increase to $\sim 60$ \%. 

The SNR models for our calculations are very similar to those 
presented by Berezhko and V\"olk \cite{berz97} except that they 
started their calculations from the free expansion stage, and 
their supernova ejecta dynamics was represented by a piston 
using the so-called thin-shell approximation.
Once the ST stage is underway, these different initial
conditions are not very important to the CR evolution.
But their numerical technique is quite different from ours,  
as mentioned in \S 1.
The results of their model with the fixed injection rate $\eta = 10^{-3}$
and Alfv\'en wave damping is similar to those of our model with Alfv\'en
wave terms included.
The overall results, that is, density compression both across the 
subshock and across the entire shock, the integrated particle spectrum,
and the total CR energy fraction are all in good agreement.
As expected, the early evolution  of our model is somewhat different, because 
we started at the beginning of the ST stage ($\hat t=1$),
while they began with the earlier free expansion stage ($\hat t=0$).

\section{Summary}

We have developed a new cosmic ray shock code in one dimensional 
spherical geometry. 
The diffusion convection equation for the CR distribution function 
is solved along with the gas dynamics equations modified to
include the diffusive CR pressure.
Simple models for thermal leakage injection and Alfv'en wave propagation
and dissipation are also included. 
In order to implement the shock tracking and AMR techniques,
we adopt a comoving spherical grid which expands with the instantaneous shock speed.
In the comoving grid, the shock remains at the same location,
so the compression rate is applied consistently to the CR distribution
at the subshock.
This results in much more accurate and efficient low energy CR acceleration and faster 
numerical convergence on coarser grid spacings, compared to the simulations
in a fixed, Eulerian grid.

We have calculated the CR acceleration during the Sedov-Taylor stage for
a typical supernova remnant in a warm interstellar medium. 
The Mach number of the initial shock is $M_s=130$.
In the simulations without the Alfv\'en wave drift and dissipation terms,  
the results are consistent with what we found for shocks with 
similar Mach numbers in plane-parallel geometry reported in
previous papers \cite{kjg02,kj05}. 
Since the diffusion length of the highest momentum ($p_{max}/mc \sim 10^6$),
$l_{diff}/r_o \sim 0.25 $, is still smaller than the shock radius,
$r_s/r_o \sim 2.9$, at $t/t_o = 10$, the pressure dilution in the
spherically expanding volume should have only minor effects. 
Although the shock expands and slows down, the CR acceleration efficiency
expressed in terms of $P_{c,2}/[\rho_0 u_{ST}^2]$ is similar to that for 
plane-parallel shocks with similar initial shock parameters.

When the drift and dissipation of Alfv'en waves in the upstream region
are included, the CR acceleration is less efficient, as expected
from previous studies \cite{mckenzi82,mark92,jon93,berz97}. 
We summarize here the main properties of the simulations with such wave 
effects included:
 
1. The postshock pressures, $P_c$ and $P_g$, normalized by
the unmodified ST shock ram pressure, $\rho_0 u_{ST}^2 \propto \hat t^{-6/5}$, 
approach time-asymptotic values quickly. 
The postshock CR pressure is about 40 \% of the ST shock ram pressure,
while the gas pressure drops from its gasdynamic value of 75 \%
to about 20 \% of the shock ram pressure.  

2. A significant shock precursor develops in response 
to nonlinear feedback from the CR 
pressure, so the subshock weakens to $M_s \approx 5$. The density
compression factor through the precursor is $\rho_1/\rho_0 \approx 2$, 
while the compression over the total transition is $\rho_2/\rho_0 \approx 8$.
However, as seen in previous spherical CR shock studies (\eg \cite{dvb95,berz97}),
the subshock does not completely disappear, as in the case of 
plane-parallel shock simulations with the self-adjusting thermal leakage 
injection model \cite{kjg02}. 

3. Both the CR momentum distribution at the shock, $g(r_s,p)=f(r_s, p)p^4$, 
and the integrated distribution, $G(p)= \int r^2 dr g(p)$, clearly
exhibit characteristic concave curvature, 
reflecting the nonlinear velocity structure in the precursor.
On the other hand, the integrated number distribution, 
$N_p(p)= \int r^2 dr f(p)p^2$ does not reveal the nonlinear features
so obviously. At nonrelativistic momenta $N_p\propto p^{-2.2}$, while
it flattens to $N_p\propto p^{-1.6}$ at the highest momenta. 

As a basic test of our new scheme for studying spherical,
CR modified shock evolution, we simulated 
SNR blast waves that were similar to SNR models
published by Berezhko \& V\"olk \cite{berz97} based on very different
numerical methods. The results are in substantial agreement.
Our new CRASH code in a comoving spherical code made
this comparison possible for the first time, because
the calculation of DSA in a comoving spherical grid 
achieves numerical convergence at a grid resolution
much coarser than that required in an Eulerian grid.

\section*{Acknowledgments}
HK was supported by the Korea Research Foundation Grant 
funded by Korea Government (MOEHRD, Basic Research Promotion Fund)
(R04-2004-000-100590)
and by KOSEF through the Astrophysical Research Center
for the Structure and Evolution of Cosmos (ARCSEC).
TWJ is supported at the University of Minnesota 
by NASA grant NNG05GF57G
and by the Minnesota Supercomputing Institute.



\begin{figure}
\begin{center}
\includegraphics*[height=40pc]{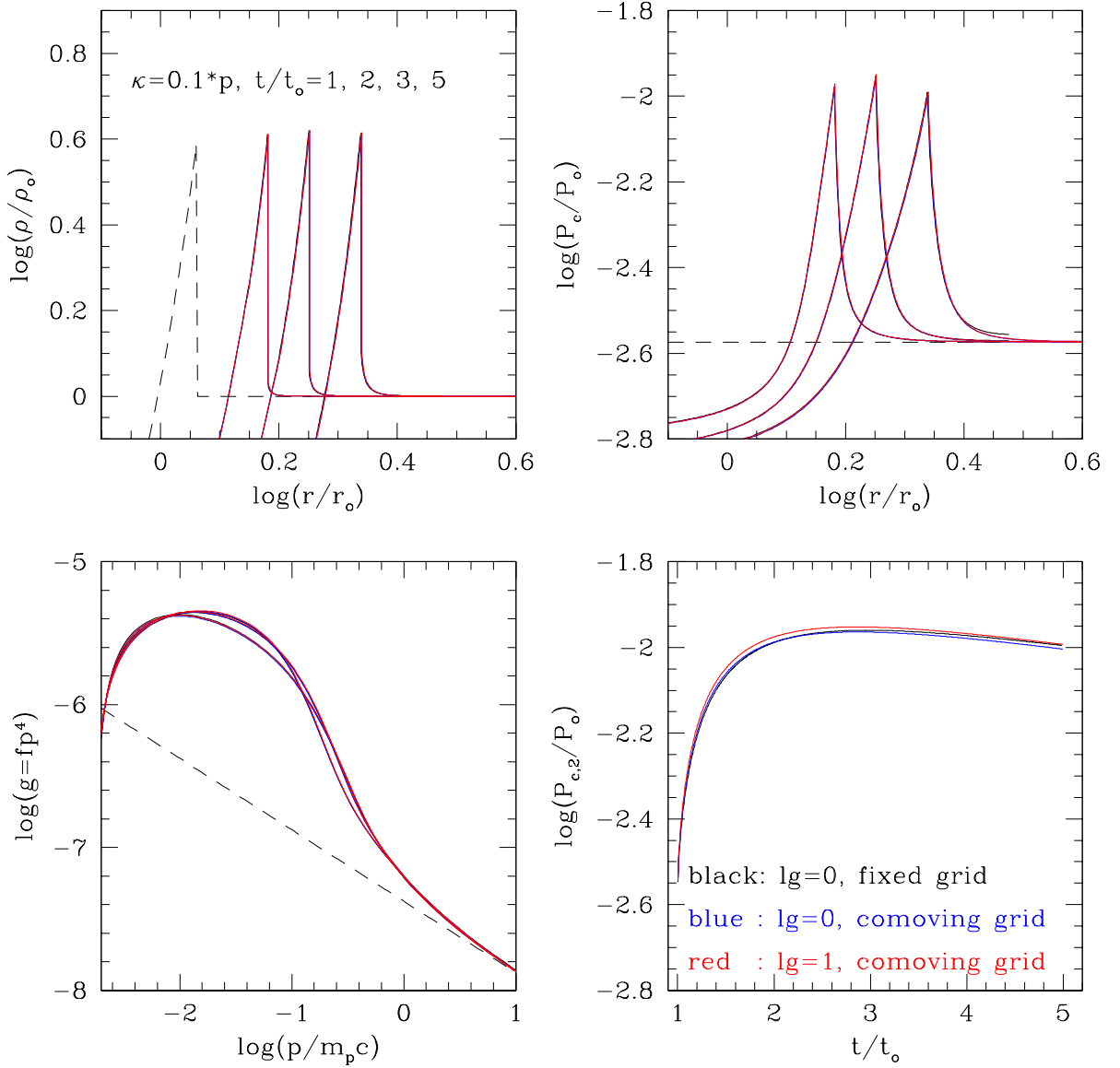}
\end{center}
\caption{
Numerical convergence test for a Sedov-Taylor blast wave in a hot ISM
with $n_H=3\times 10^{-3}{\rm cm}^{-3}$ and $T=10^6$K. 
The initial Mach number of the shock is $M_s=13$. 
A preexisting CR population, $f(p)\propto p^{-4.5}$, corresponding
to $P_{c,0}= 4 P_{g,0}$, is assumed.
The diffusion model is $\tilde\kappa(p)=0.1p(\rho_0/\rho)$ in code units.
The black dashed lines show the initial conditions. 
The CR spectrum at the shock , $g(r_s,p) = f(r_s,p)p^4 $, is shown in the 
lower left panel, while the postshock CR pressure normalized
by the fiducial pressure is shown as a
function of time in the lower right panel.
Black lines are for the calculation in the fixed Eulerian grid
without AMR,
blue lines for the calculation in the comoving grid without AMR,
and red lines for the calculation in the comoving grid with one
level of AMR.
}

\label{converg}
\end{figure}

\begin{figure}
\begin{center}
\includegraphics*[height=40pc]{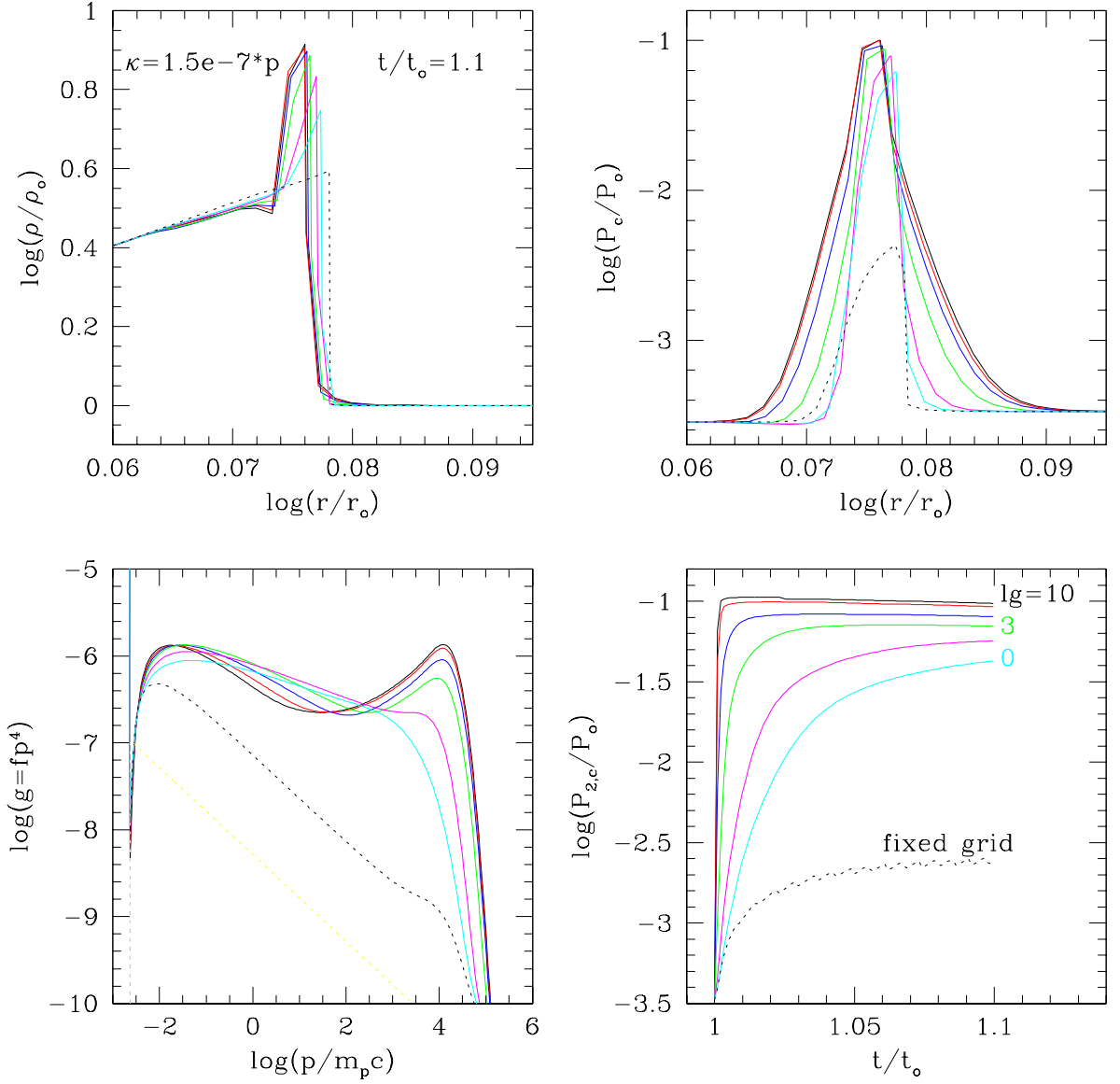}
\end{center}
\caption{
Numerical convergence test for a Sedov-Taylor blast wave in a hot ISM
with $n_H=3\times10^{-3}{\rm cm}^{-3}$ and $T=10^6$K. 
A preexisting CR population, $f(p)\propto p^{-4.5}$, corresponding
to $P_{c,0}=0.5 P_{g,0}$, is assumed (yellow dotted line in the low right panel).
The diffusion model is $\tilde \kappa(p)=1.5\times 10^{-7} p(\rho_0/\rho)$.
The CR spectrum at the shock, $g(r_s,p) = f(r_s,p)p^4 $, is shown in the
lower left panel, while the postshock CR pressure is shown as a
function of time in the lower right panel.
Black dotted lines are used for the calculation with the fixed Eulerian
grid without AMR.
For the calculations with the comoving grid, different colored lines are
used for different levels of AMR:
cyan, magenta, green, blue, red and black for
$l_g=0$, 1, 3, 5, 8, and 10, respectively.
}
\label{converg2}
\end{figure}

\begin{figure}
\begin{center}
\includegraphics*[height=40pc]{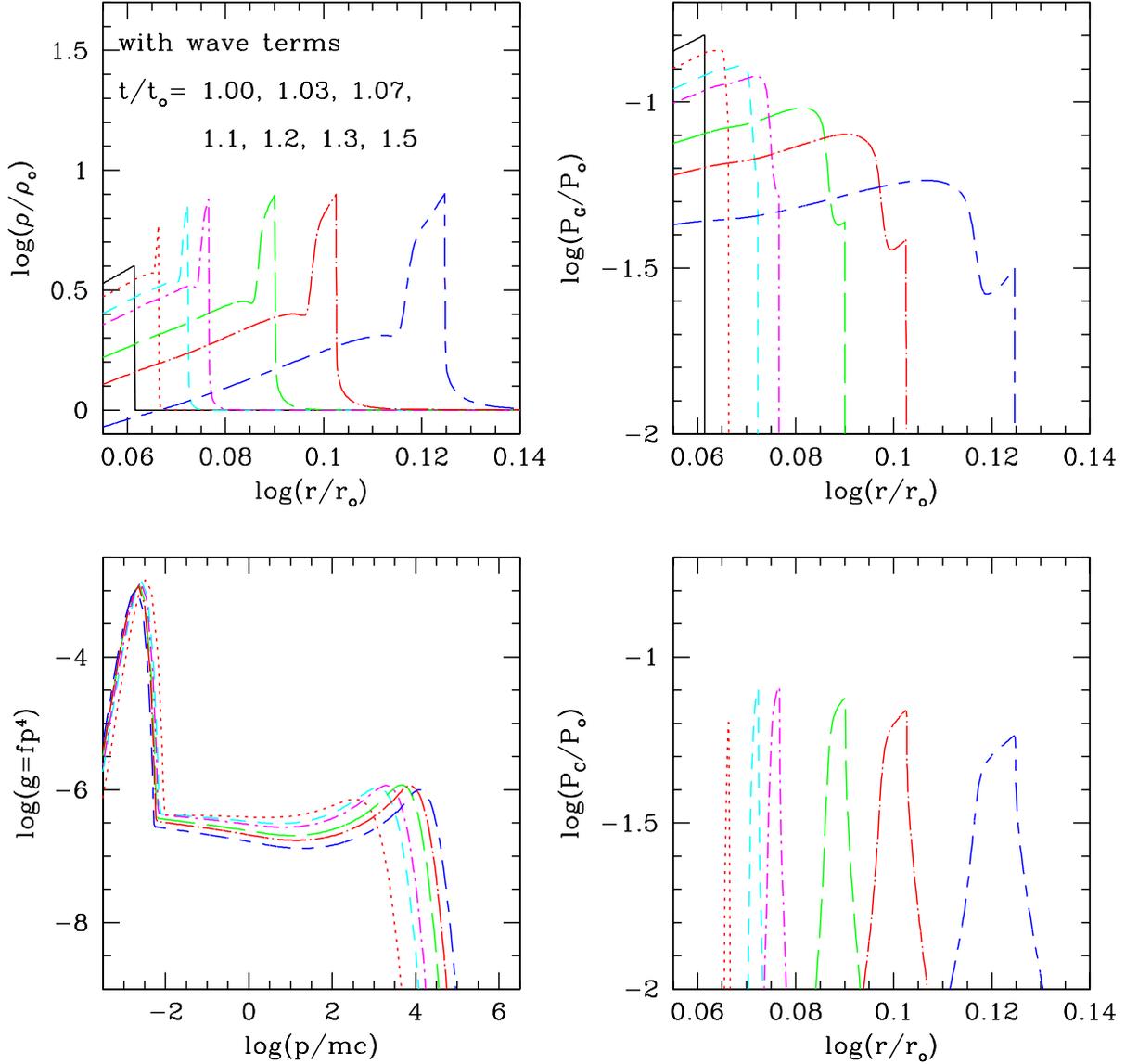}
\end{center}
\caption{
Evolution a SNR expanding into the warm uniform ISM
with $n_H=0.3 {\rm cm}^{-3}$ and $T=10^4$K.
The model parameters are
$E_o=10^{51}$ ergs, $M_{sn}=10 M_{Sun}$, and $B_{\mu}=5$.
The model assumes a preexisting CR population of $f(p)\propto p^{-4.5}$,
with $P_{c,0}=0.5P_{g,0}$.
The injection parameter for thermal leakage injection, $\epsilon_B=0.16$.
The lower left panel shows the CR spectrum at the shock,
$g(r_s, p) = f(r_s,p)p^4 $.
The initial condition at $ \hat t=1.0$ (solid line) is set by the
Sedov-Taylor similarity solution.
}
\end{figure}

\begin{figure}
\begin{center}
\includegraphics*[height=40pc]{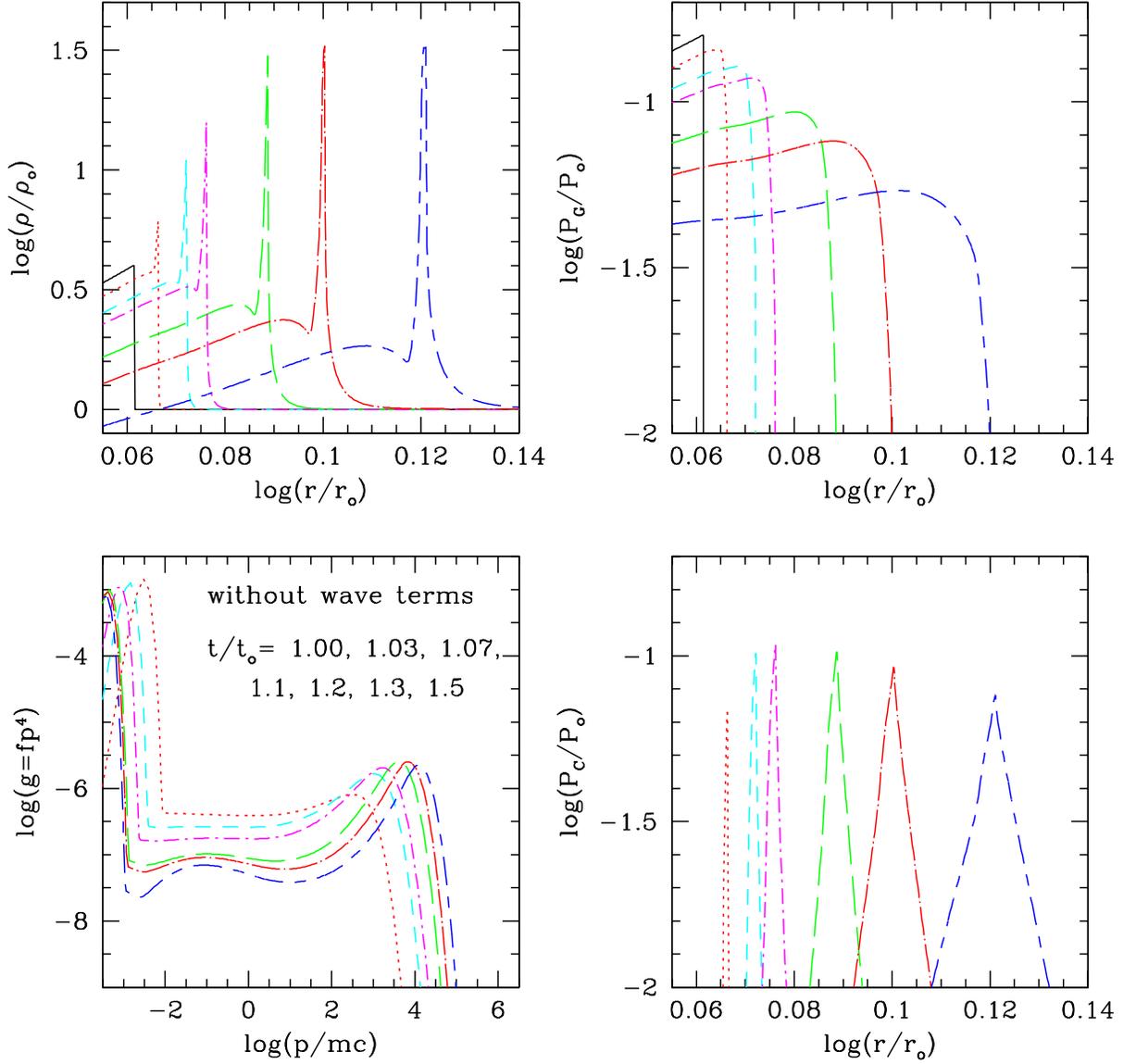}
\end{center}
\caption{
Same as Fig. 2 except the Alfv\'en wave drift and heating terms are dropped.
}
\end{figure}

\begin{figure}
\begin{center}
\includegraphics*[height=40pc]{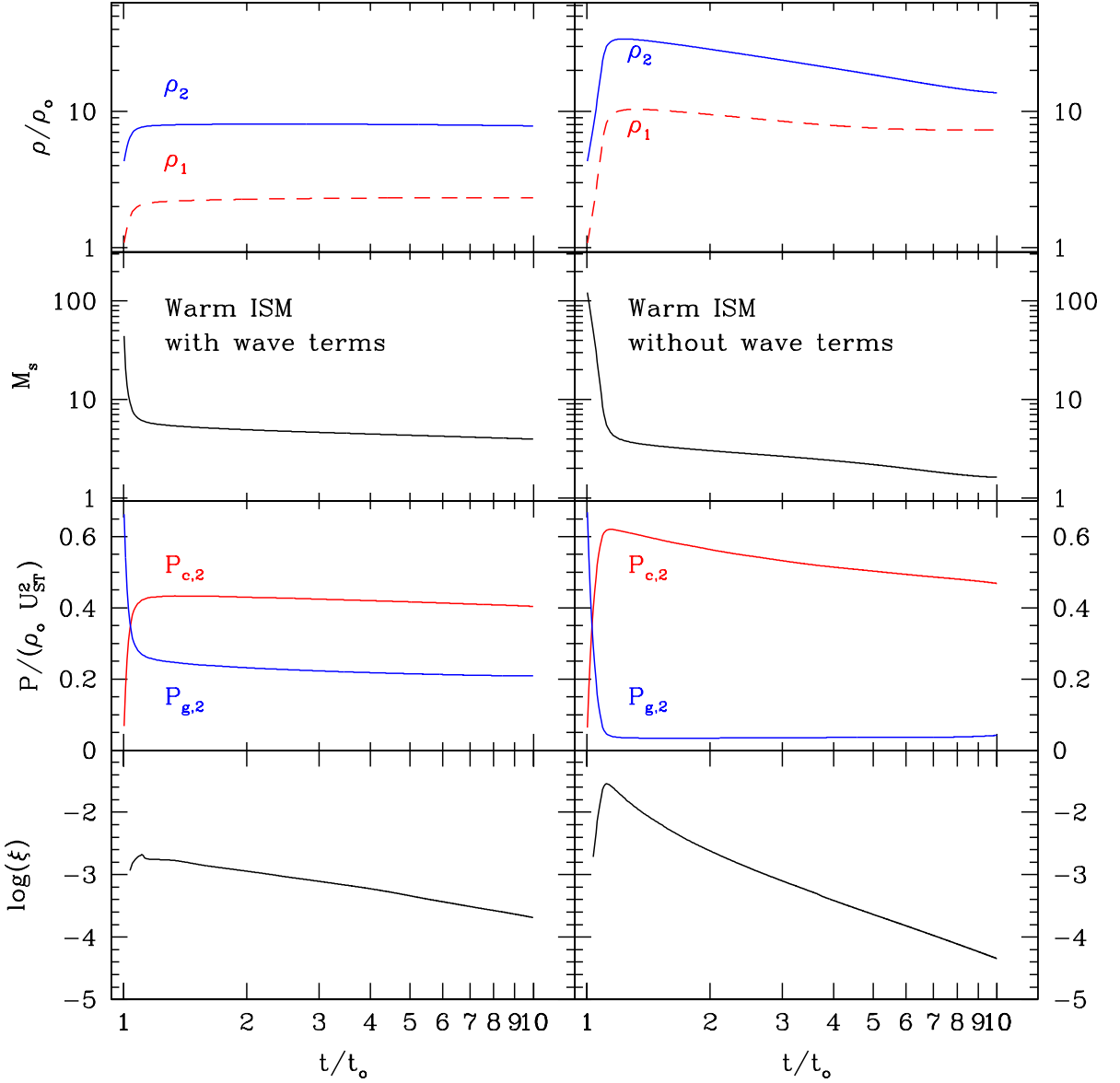}
\end{center}
\caption{
Pre-subshock density, $\rho_1$, post-subshock density,
$\rho_2$, shock Mach number, $M_s$, post-subshock CR and gas pressure
in units of the ram pressure of the unmodified Sedov-Taylor solution, 
$\rho_0 U_{ST}^2 \propto \hat t^{-6/5}$, and the CR injection parameter, $\xi$, are plotted.
}
\end{figure}

\begin{figure}
\begin{center}
\includegraphics*[height=40pc]{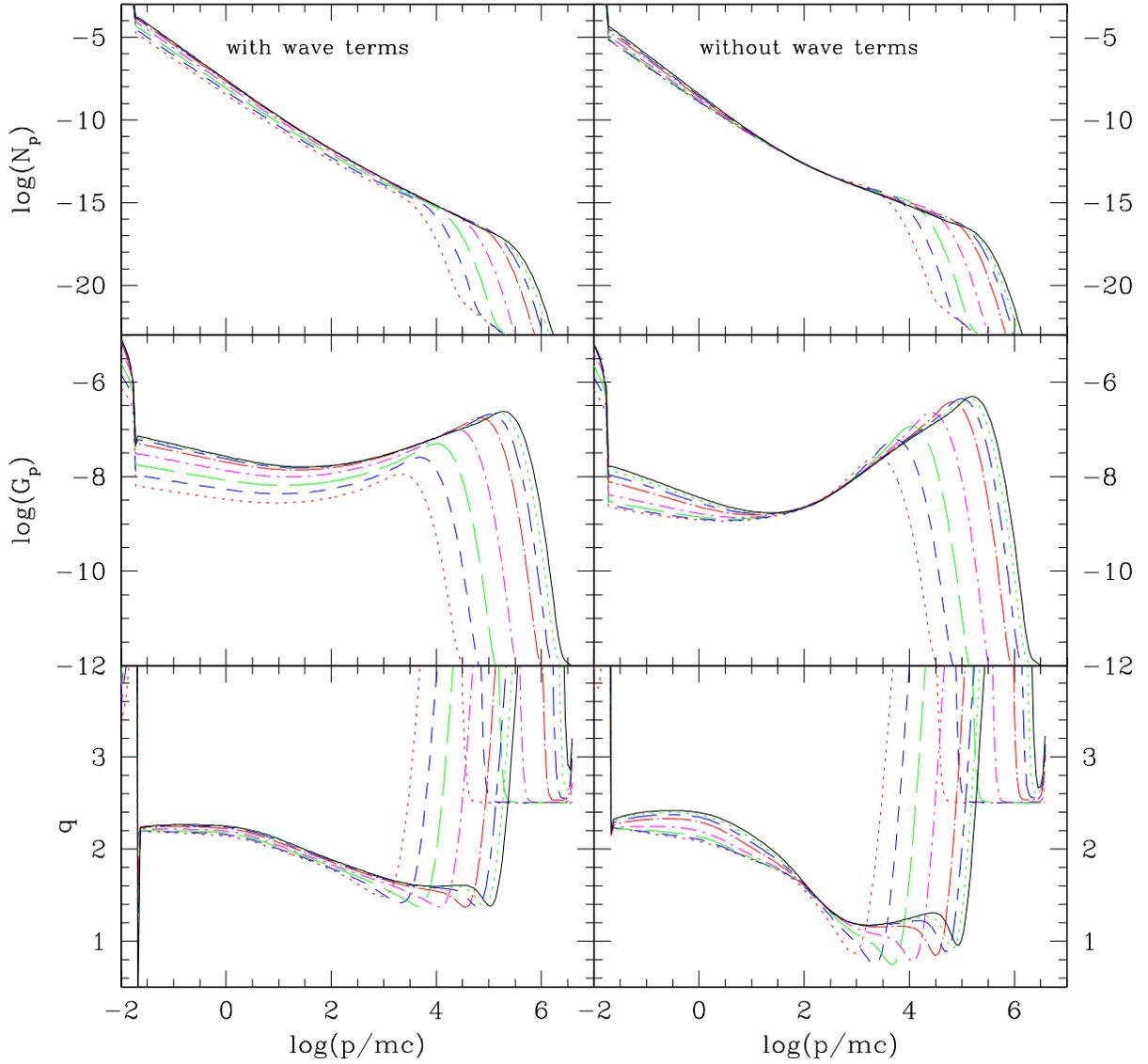}
\end{center}
\caption{
Integrated CR number, $N_p=  \int r^2 dr  f(p) p^2 $, and
$G_p= N_p p^2 $ in arbitrary units and 
the slope, $q = - d (\ln N_p)/ d \ln p$, are shown 
at $\hat t=$ 1.1, 1.2, 1.4, 2., 4., 6., 8., and 10.
}
\end{figure}

\begin{figure}
\begin{center}
\includegraphics*[height=40pc]{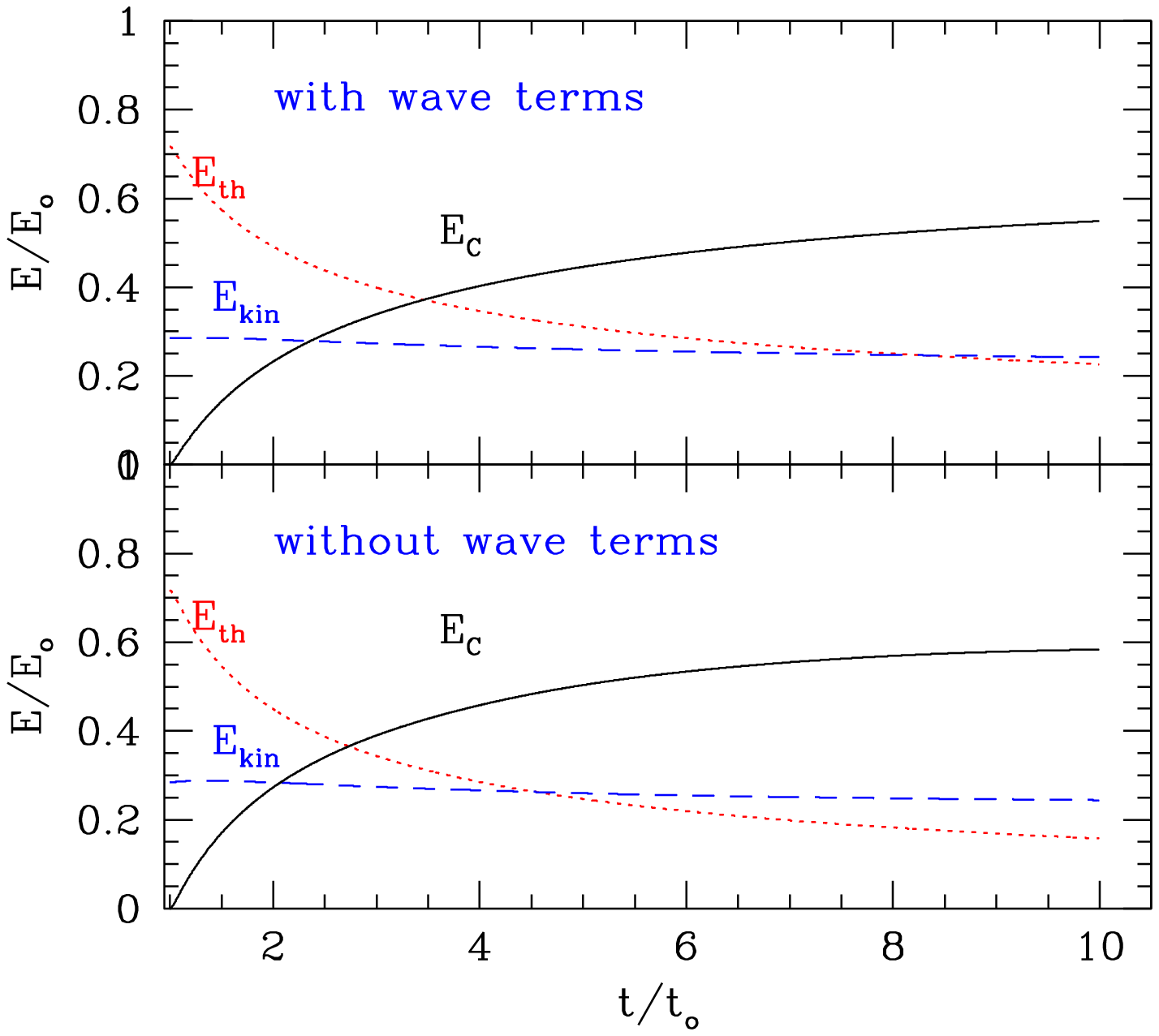}
\end{center}
\caption{
Integrated thermal, kinetic and CR energies inside the simulation volume
as a function of time for the models shown in Figs. 3 and 4.
}
\end{figure}


\begin{thebibliography}{00}




\bibitem{ach82}
Achterberg, A. Astron. \& Astrophys. 98 (1982) 195

\bibitem{achbl86}
Achterberg, A. \& Blandford, R. D. M.N.R.A.S. 218 (1986) 551

\bibitem{bell78}
Bell, A. R. M.N.R.A.S. 182 (1978) 147

\bibitem{berel99}
Berezhko E. G. \& Ellison, D.C. Astrophys. J. 526 (1999) 385

\bibitem{berz94} 
Berezhko E.G., Yelshin V.K., \& Ksenofontov L.T. Astroparticle Physics 
2 (1994) 215

\bibitem{berz95}
Berezhko E.G., Ksenofontov L.T., \& Yelshin V.K. Nuclear Physics B 39a (1995) 171

\bibitem{berz97} 
Berezhko E.G., V\"olk, H.J. Astroparticle Physics 7  (1997) 183 

\bibitem{berz00} 
Berezhko E.G., V\"olk, H.J. Astron. \& Astrophys. 357  (2000)  283 

\bibitem{blas02}
Blasi, P., Astroparticle Physics 16 (2002) 429

\bibitem{chev95}
Chevalier, R. A. \& Blondin, J. M., Astrophys. J. 444 (1995), 312

\bibitem{dru83} 
Drury, L.~O'C. Rept. Prog. Phys. 46 (1983) 973

\bibitem{dvb95}
Drury, L. O'C., V\"olk, H. J. \& Berezhko, E. G. Astron. \& Astrophys. 299 (1995) 299

\bibitem{drury00} 
Drury, L.~O'C. \& Mendonca, J.~T. Physics of Plasmas, 7 (2000) 5148 

\bibitem{elber99}
Ellison, D. C. \& Berezhko, E. G. 26th ICRC 4 (1999) 446

\bibitem{ellison05} 
Ellison, D. C., Decourchelle, A., Ballet, J. Astron. \& Astrophys. 429 (2005), 569

\bibitem{gies00}
Gieseler, U. D. J., Jones, T. W. \& Kang, H. Astron. \& Astrophys. 364 (2000) 911

\bibitem{jon93}
Jones, T.~W. Astrophys. J. 619 (1993) 619

\bibitem{jk05}
Jones, T. W. \& Kang, H. Astroparticle Phys. 24 (2005) 75

\bibitem{jun96}
Jun, B.-I. \& Norman, M. L. Astrophys. J. 465 (1996) 800

\bibitem{kj91}
Kang, H. \& Jones, T. W., M.N.R.A.S. 249 (1991) 439

\bibitem{kjr92} 
Kang, H., Jones, T.~W., and Ryu, D. Astrophys. J. 385 (1992) 193

\bibitem{kj02}
Kang, H. \& Jones, T. W., Journal of Korean Astronomical Society, 35 (2002), 159 
\bibitem{kjls01} 
Kang, H., Jones, T. W., LeVeque, R. J., Shyue, K. M. Astrophys. J. 550 
(2001) 737

\bibitem{kjg02} 
Kang, H., Jones, T. W., \& Gieseler, U.D.J. Astrophys. J. 579 (2002) 337

\bibitem{kj05} 
Kang, H., \& Jones, T. W. Astrophys. J. 620 (2005) 44 


\bibitem{mal98} 
Malkov M.A. Phys. Rev. E,  58 (1998) 4911

\bibitem{malvol98} 
Malkov, M.A., and V\"olk, H.J. Adv. Space Res. 21 (1998) 551

\bibitem{maldru01} 
Malkov, M.A., \& Drury, L.O'C. Rep. Progr. Phys. 64 (2001) 429

\bibitem{mark92}
Markiewicz, W. J., Drury, L. O'C. \& V\"olk, H. J. Astron. \& Astrophys. 236 (1992) 487

\bibitem{martel98}
Martel, H. \& Shapiro, P.~R. M.N.R.A.S. 297 (1998) 467

\bibitem{mckenzi82}
McKenzie, J. F., \& V\"olk, H. J., Astron. \& Astrophys. 116 (1982) 191  

\bibitem{ptus03}
Ptuskin, V. S. \& Zirakashvili, V. N. Astron. \& Astrophys. 403 (2003) 1

\bibitem{ryu93} Ryu, D., Ostriker, J. P., Kang, H., \& Cen, R.
ApJ, 414 (1993) 1 

\bibitem{skill75}
Skilling, J., M.N.R.A.S., 223 (1975) 353

\end{thebibliography}
\end{document}